%% file: main.tex
\PassOptionsToPackage{dvipsnames}{xcolor}

\documentclass[acmsmall,screen,nonacm,natbib=false]{acmart}

\input{math_commands.tex}

\usepackage{url}
\usepackage{etoc}

\usepackage{common} 
\usepackage{model_names}

\newcommand{\lines}{\text{lines}}

\newcommand{\vPass}{\text{pass}}

\newcommand{\vAccepted}{\text{accepted}}
\newcommand{\vRejected}{\text{rejected}}

\newcommand{\Bare}{\texttt{bare}\xspace}
\newcommand{\Acr}{\texttt{acr}\xspace}
\newcommand{\AgInt}{\textsc{AgInt}\xspace}
\newcommand{\AcrBody}{\textsc{AgInt}@\texttt{acr}\xspace}

\newcommand{\AgIntRef}{\AgInt(ref)}

\DeclareRobustCommand{\metricname}[1]{\ifmmode\text{#1}\else#1\xspace\fi}
\newcommand{\MRR}{\metricname{MRR}}
\newcommand{\WastedAt}[1]{\metricname{wasted@#1}}
\newcommand{\AllAt}[1]{\metricname{All@#1}}

\newcommand{\WastedAtN}{\WastedAt{\ensuremath{N}}}
\newcommand{\AllAtN}{\AllAt{\ensuremath{N}}}

\newcommand{\FoneReject}{\metricname{F1}}
\newcommand{\FoneOf}[1]{\metricname{F1(#1)}}
\newcommand{\FoneAccept}{\FoneOf{accept}}

\newcommand{\MacroFone}{\metricname{Macro-F1}}
\newcommand{\Accuracy}{\metricname{Accuracy}}

\newcommand{\RecallReject}{\metricname{recall}}

\newcommand{\SBL}{SWE-bench-Live\xspace}
\newcommand{\DSW}{DeepSWE\xspace}

\newcommand{\Defects}{Defects4J\xspace}
\newcommand{\GitBug}{GitBug-Java\xspace}

\newcommand{\SplitLarge}{\textsc{large100}\xspace}

\usepackage[
    backend=biber,
    style=authoryear,
    natbib=true,
    maxcitenames=2,
    maxbibnames=5,
    sortcites=true,
    uniquename=false,
    uniquelist=false
]{biblatex}
\DeclareSourcemap{
  \maps[datatype=bibtex]{
    \map{\step[fieldsource=doi, final]\step[fieldset=url, null]}
    \map{\step[fieldsource=eprint, final]\step[fieldset=url, null]}
    \map{\step[fieldset=isbn, null]\step[fieldset=issn, null]}
  }
}

\graphicspath{{figures/}}

\title{Agentic Interpretation: Harness-Governed Evidence Construction for Software Agents}

\author{Yifeng He}
\email{yfhe.cs@gmail.com}
\authornote{This work was done during an internship at AWS.}
\affiliation{%
  \institution{University of California, Davis}
  \city{Davis}
  \state{California}
  \country{USA}
}
\author{Yanjun Wang}
\email{yanjunw@amazon.com}
\affiliation{%
  \institution{AWS AI Labs}
  \country{USA}
}
\author{Siwei Cui}
\email{siweicui@amazon.com}
\affiliation{%
  \institution{AWS AI Labs}
  \country{USA}
}
\author{Aritra Sengupta}
\email{aritras@amazon.com}
\affiliation{%
  \institution{AWS AI Labs}
  \country{USA}
}
\author{Murali Krishna Ramanathan}
\email{mkraman@amazon.com}
\affiliation{%
  \institution{AWS AI Labs}
  \country{USA}
}
\renewcommand{\shortauthors}{He et al.}

\begin{document}

\begin{abstract}
\input{src/abstract}
\end{abstract}

\maketitle

\input{src/intro}
\input{src/background}
\input{src/method}
\input{src/experiments}
\input{src/conclusion}
\input{src/data_availability}


\printbibliography


\end{document}

%% file: math_commands.tex
\usepackage{amsmath,amsfonts,bm}

\newcommand{\newterm}[1]{\emph{#1}}

\def\eqref#1{equation~\ref{#1}}

\def\1{\bm{1}}

\DeclareMathAlphabet{\mathsfit}{\encodingdefault}{\sfdefault}{m}{sl}
\SetMathAlphabet{\mathsfit}{bold}{\encodingdefault}{\sfdefault}{bx}{n}

%% file: src/abstract.tex
Software agents answer behavioral questions about code, such as whether
a patch will pass its tests or which lines caused an observed failure.
These tasks require gathering and combining evidence across a repository, yet
most agent scaffolds leave intermediate findings in a free-form trajectory,
making them difficult to reuse and easy to disconnect from the final answer.
We present \emph{Agentic Interpretation} (\AgInt), a framework for
harness-governed evidence construction. The model proposes typed records,
while a deterministic harness governs their admission, grading, and
persistence, reports unresolved requirements, and licenses an answer under
task-specific rules. The resulting certificate gives the model persistent
external state for subsequent reasoning and leaves an auditable record of the
decision. Formally, append-only certificate states form a grounding lattice
ordered by record inclusion, and each admitted record extends the state by
union; the model still chooses what to investigate and when to propose an
answer.

We evaluate \AgInt{} with four models on \SBL{} and \DSW{} for patch
acceptance and \Defects{} and \GitBug{} for fault localization. \AgInt{} leads
the plain agent and Agentic Code Reasoning on every reported patch-acceptance
metric and generally ranks responsible lines higher while inspecting fewer
irrelevant lines in fault localization. For \GLM{}, reject-class F1 rises from
\num{0.256} to \num{0.543} on \SBL{}, \AllAt{1} rises from \num{0.140} to
\num{0.400} on \Defects{}, and wasted inspected lines fall from \num{31.5} to
\num{11.7} on \GitBug{}. A matched-template control separates the mechanism:
the persistent scaffold carries most of the fault-localization gain, while the
certificate body and gate drive patch-acceptance calibration. With execution
licensed, \AgInt{} produces harness-reproduced witnesses for \num{293} of
\num{475} null-dereference candidates versus \num{253} for a matched shell
agent, and \num{62} of \num{117} resource-leak candidates versus \num{58}.
Across the reported efficiency measurements, it uses \num{35}\% fewer
executions and reduces cost per candidate by \num{59}\%.

%% file: src/intro.tex
\section{Introduction}
\label{sec:intro}

Interpreting program behavior, rather than merely retrieving or
modifying code, underlies change review, fault localization, and warning
triage. Software
agents increasingly answer the same questions on our behalf, and anyone
who acts on an agent's verdict must judge it by the evidence the agent
gathered. That evidence rarely sits in one place, since the path that decides
a question may span tests, callbacks, libraries, configuration, and
several files. A fixed query rarely traces this path; a language-model agent
can choose sources to inspect, form a hypothesis, and revise it as evidence
arrives. For these tasks, reliable reasoning requires not only proposing a
conclusion but retaining the evidence gathered along the way.

Language-model agents extend earlier reasoning methods by gathering
evidence from an external environment. Chain-of-thought, self-consistency, and
thought search organize or explore intermediate model text
\citep{wei2022cot,wang2023selfconsistency,yao2023tot}, but they gather
no new evidence. An agent instead proposes
a tool call, and a harness executes it and returns an observation
\citep{yao2023react,yang2024sweagent,luo2025agentsurvey,anthropic2026harness}.
The trajectory is therefore no longer a generated explanation but an
accumulated record of claims, actions, and observations. Because the
harness mediates every action and records its effect, that record opens an
auditable surface.

Current software agents nevertheless organize gathered evidence as
a free-form trajectory: the model decides what to claim, how observations
support those claims, and when to stop
\citep{yao2023react,yang2024sweagent,wu2025agenticreasoning}. A harness may
constrain tool calls, preserve the transcript, or check individual steps, but
without shared state and transition rules it does not define which records are
admissible, how their evidence composes, or when the record licenses a verdict.
Agentic Code Reasoning (\Acr) introduces a task-specific semi-formal certificate
template, but leaves its construction semantics open \citep{ugare2026acr}.
We argue that the harness, not the model, should set these rules, and
that two rules suffice. We call this harness design \emph{Agentic
Interpretation} (\AgInt). The model proposes typed records for a task-specific
certificate. \textsc{Step} admits a proposal only when deterministic checks
accept it, grades its evidence, and appends it to the certificate.
\textsc{Close} returns an answer only when the certificate meets the
obligations that answer owes. Admitted evidence, rather than model thoughts,
thus becomes the agent's state, and the model's prose can neither grade its
own evidence nor license its own verdict. After each step, the harness returns
the certificate and its unresolved obligations, so the model builds its next
proposal on durable state instead of recovering its investigation from a
transcript. The certificate also outlives the session, so a third party can
recheck the verdict without rerunning the model. Because both rules are
deterministic harness code, we can also state and check properties of the
certificate independently of the stochastic model policy, as operational
semantics and program-analysis frameworks do for programs
\citep{cousot1979systematic}.

These two rules also determine the harness-visible state space. Because
\textsc{Step} only appends immutable records, the pre-close states are sets of
records ordered by inclusion. They form a powerset lattice, which we call the
\emph{grounding lattice}, and each admitted record extends the state by union.
This structure adapts two ideas from abstract interpretation to agent reasoning:
ordered, monotone state and a worklist-like frontier of unresolved obligations
\citep{cousot1977abstract,cousot1979systematic}. The harness reports that
frontier and joins admitted records into persistent state, while the model
chooses what to investigate, which record to propose, and when to attempt a
verdict. \AgInt{} thereby brings the state organization of program analysis
into an agent harness while retaining model control over search and answer
timing.

To isolate the gain that the scaffold itself delivers, we first
evaluate \AgInt without code execution, on patch acceptance
\citep{xiong2018identifying,wang2020apca} and fault localization
\citep{wong2016survey,kang2024autofl}, and then on execution-enabled bug witnessing.
Across four models and four labelled datasets, \AgInt raises \GLM's patch
accuracy by eight points and reject-class F1 from \num{0.256} to \num{0.543} on
\SBL; it raises fault-localization MRR from \num{0.323} to \num{0.633} on
\Defects and cuts wasted inspected lines from \num{31.5} to \num{11.7} on
\GitBug. The gains are largest where the plain agent is weakest: on \SBL, \Qwen's reject-class
F1 rises from \num{0.032} to \num{0.496}. A matched-template ablation preserves
\AgInt's scaffold while substituting \Acr's unchanged certificate body,
separating the scaffold's persistence and progress guidance from the redesigned
body and gate. In bug witnessing, every positive carries a concrete witness
reproduced by the harness. Against a matched shell-agent baseline, \AgInt
witnesses \num{293} of \num{475} pooled null-dereference candidates versus
\num{253}, and \num{62} of \num{117} resource-leak candidates versus \num{58}.
Our contributions are:
\begin{itemize}[noitemsep,topsep=2pt,leftmargin=*]
\item A simple, principled harness design for agent evidence
      collection, in which two rules, \textsc{Step} and \textsc{Close}, move
      admission, evidence grading, and verdict licensing out of the model and
      into deterministic harness code (\autoref{sec:method}).
\item A grounding-lattice semantics for append-only evidence state, together
      with an obligation-guided agent loop that adapts worklist organization
      to model-directed investigation
      (\autoref{sec:method:lattice}, \autoref{sec:method:loop}).
\item Structural guarantees that follow from this design (bounded
      grade ascent, order-independent rechecking of a fixed certificate,
      and harness-reproduced witnesses), with a replay audit of rechecking
      on real trajectories (\autoref{sec:method:join}).
\item Controlled patch-acceptance and fault-localization results,
      including a matched-template ablation
      (\autoref{sec:experiments:reasoning}).
\item A bug-witnessing evaluation in which every positive is backed
      by a harness-reproduced input (\autoref{sec:experiments:witness}).
\end{itemize}

%% file: src/background.tex
\section{Background and Related Work}
\label{sec:background}

\subsection{Agents, Harnesses, and Certificates}
\label{sec:background:agents}

An LLM agent pairs a stochastic model that proposes actions with a
deterministic harness that supplies tools and executes or refuses those actions
\citep{luo2025agentsurvey,anthropic2026harness}. We use \emph{model} for
the stochastic component, \emph{harness} for the deterministic code, and
\emph{agent} for the pair. This distinction is the paper's
trust boundary: admission, evidence grading, and verdict licensing are harness
decisions.

A \newterm{certificate} is the durable record of claims and evidence
behind an answer; its template fixes the task questions and supported answer
branches. \Acr introduced task-specific certificates for code reasoning
\citep{ugare2026acr}. \AgInt retains this representation and adds deterministic
transition semantics for constructing and licensing it. A matched-template
control separates this machinery from the certificate body
(\autoref{sec:experiments:control}).

\subsection{Ordered Domains, Fixed Points, and Abstract Interpretation}
\label{sec:background:lattices}

A \newterm{partial order} $(L,\sqsubseteq)$ orders states by
information: $x\sqsubseteq y$ means that $y$ contains at least as much
information as $x$. A \newterm{lattice} provides a least upper bound, or
\newterm{join}, $x\sqcup y$. In the powerset lattice
$(\mathcal{P}(A),\subseteq)$ used later, join is union and the least element is
$\emptyset$. A function $F$ is \newterm{monotone} when
$x\sqsubseteq y$ implies $F(x)\sqsubseteq F(y)$; repeated monotone updates form an ascending chain.
A \newterm{fixed point} is a state that the update leaves unchanged, and
finite height ensures that an ascending iteration reaches one after finitely
many strict increases. A \newterm{worklist algorithm} processes a pending item, joins new
information into the current state, and re-enqueues work affected by that
change.
It reaches a fixed point when no pending update can change the state. Abstract interpretation
uses these structures to summarize concrete program states, join information,
and iterate transfer functions to a fixed point
\citep{cousot1977abstract,cousot1979systematic}.
A finite-height domain terminates by this ascent argument. An
infinite-height domain instead needs a \newterm{widening}, which jumps ahead
in the order at the cost of precision, or a coarser finite-height view whose
ascent is bounded instead.
The approximation direction determines what an analysis establishes. A
\newterm{may} analysis overapproximates behaviors, so an omitted behavior is
absent but a reported one may be spurious. A \newterm{must} analysis
underapproximates them, so every reported behavior can occur but an omitted
one may still occur \citep{godefroid2010maymust,ball2005falsification}.

\subsection{Related Work}
\label{sec:related}

\paragraph{Agentic reasoning}
Agentic-reasoning and thought-search methods govern how a model
plans, calls tools, and revises its trajectory
\citep{yao2023react,wei2026agenticsurvey,wei2022cot,
wang2023selfconsistency,yao2023tot,besta2024got}. \AgInt instead governs the
meaning of the accumulated evidence record. Rigor gates and step checkers
constrain individual inferences
\citep{kon2025curie,poesia2024logicguide,ling2023naturalprogram}, but do not by
themselves define when the accumulated record licenses a verdict.

\paragraph{Execution-free code judgment}
CRUXEval and REval expose weaknesses in execution reasoning
\citep{gu2024cruxeval,chen2025reval}, while SWE-bench establishes
repository-scale patch tasks \citep{jimenez2024swebench}. We evaluate newer
\SBL and \DSW records \citep{li2026repolaunch,datacurve2026deepswe}.
Dockerless gathers repository evidence without execution but leaves the verdict
to an LLM \citep{zeng2026dockerless}; \AgInt uses deterministic licensing.
For fault localization, LLMAO and AutoFL rank locations
\citep{yang2024llmao,kang2024autofl}, whereas \AgInt binds each bounded region
to its licensing evidence.

\paragraph{Producer--checker trust boundaries}
Proof-carrying code separates an untrusted producer from a small certificate
checker \citep{necula1997pcc}. Proof-producing translation, Lemur, and Baldur
apply this boundary to generated programs or proofs
\citep{parthasarathy2024trustworthy,wu2024lemur,first2023baldur}, and their
checkers discharge semantic obligations in a formal logic. \AgInt draws a
graded boundary instead. Witnessed positives recover a one-sided semantic
guarantee, whereas cited records establish only that a reference resolves. A
concurrent pipeline calculus also combines a proof-carrying boundary, graded
grounding, and an emission gate \citep{koomullil2026emission}. Relative to that calculus, \AgInt{} focuses on sequential certificate
construction with deterministic admission, append-only updates, task-specific
obligations, and order-independent rechecking of a fixed certificate.

\paragraph{Violation witnesses}
SV-COMP standardizes violation witnesses that independent validators check
\citep{beyer2022witnesses}, and LLift uses an LLM to adjudicate kernel warnings
that symbolic analysis leaves unresolved \citep{li2024llift}. \AgInt places
the model upstream of a similar endpoint. Only an observation that the harness
reproduces under a declared oracle can license a witnessed positive, and the
certificate records how the agent reached the flagged site.

%% file: src/method.tex
\section{The Grounding Calculus of Agentic Interpretation}
\label{sec:method}

In \AgInt, the model reasons by proposing typed records for a
task-specific certificate. The harness admits or refuses proposals, persists
admitted records, reports unresolved requirements, and licenses an answer
under task-specific rules. The admitted certificate becomes durable state for
the model's subsequent proposals.

\begin{definition}[Atoms and certificate state]
An \newterm{atom} is one immutable, normalized record the harness
appends. During construction, the pre-close state $E$ is the finite
set of admitted nonterminal atoms. If \textsc{Close} licenses $v$, the harness
stores the sealed certificate
$\mathsf{seal}(E,v)=E\cup\{\langle v\rangle\}$; that terminal record has no
successor state. For a fixed task and evidence environment, grading,
resolution, obligations, and licensing are deterministic functions of $E$.
Admission additionally reads the proposed move. None of these checks reads
hidden model state or the free-form transcript.
\end{definition}

Authority over this state is split: the untrusted model may propose
a record, but the deterministic harness alone admits it, assigns its evidence
grade, and decides whether the accumulated certificate licenses an answer.
The model directs the investigation, but its prose cannot assign its own grade
or authorize a verdict.

\begin{definition}[Grounding task]
A \newterm{grounding task} states what the agent works with, what it must
produce, and what an answer must show. It fixes an immutable evidence
environment $\Gamma$, typically a read-only repository checkout together with
the patch or tests in the task record. It fixes the shape of the output:
the answers the agent may return and a certificate template with a finite set
of \newterm{slots}, whose evidence is scored on a chain of \newterm{grades}.
It fixes the requirements: for each answer, which slots must reach which
grades. The harness decides whether to admit each proposed record
($\mathsf{adm}$), computes the resulting grades ($\mathsf{grade}$), and licenses
an answer ($\mathsf{lic}$) only when its grade requirements ($\mathsf{req}$) are
met, no decisive outcome is contested, and the settled outcomes support it.
\end{definition}

Grades have task-defined meanings. In our tasks, \emph{cited}
means that a source reference resolves; it does not show that the surrounding
interpretation is correct. Licensing can additionally check relationships
among typed fields and records, and because the requirements read the
certificate, what an answer owes can depend on what the record already says.

A task is \newterm{well formed} when extending $E$ cannot remove an
available slot or lower a grade. This condition governs progress, not
licensing: a new atom may expose another obligation or create a tie. Tasks
with sub-slot expansion fix a bound $B_{\mathcal{T}}$ on how many sub-slots a run may open, keeping the
available slot set finite and recoverable from $E$; $B_{\mathcal{T}}=0$
removes the expansion move.

Over a fixed task, environment, and state, the harness computes four
judgments. Admission $\mathcal{T};\Gamma;E \vdash p \leadsto a$ holds when
$\mathsf{adm}_{\mathcal{T}}$ admits proposal $p$ as atom $a$; grading
$\mathcal{T};\Gamma;E \vdash s : g$ holds when
$\mathsf{grade}_{\mathcal{T}}(\Gamma,E,s)=g$. Two inference rules derive
transition and closure:
\[
\frac{\mathcal{T};\Gamma;E\vdash p\leadsto a}
     {\mathcal{T};\Gamma\vdash E\xrightarrow{p} E\cup\{a\}}
\;(\textsc{Step})
\qquad
\frac{\mathsf{lic}_{\mathcal{T}}(\Gamma,E,v)}
     {\mathcal{T};\Gamma\vdash E\Downarrow v}
\;(\textsc{Close}).
\]
\textsc{Step} appends one admitted atom. A commit proposal invokes
\textsc{Close}, which reads the pre-close state $E$ and, when it succeeds,
stores $\mathsf{seal}(E,v)$; the commit atom contributes no grade and is not a
premise of its own licensing. A refusal has no constructive \textsc{Step} and
leaves $E$ unchanged. The model chooses proposals but never decides whether
either rule applies.

\subsection{Certificate State as a Grounding Lattice}
\label{sec:method:lattice}

Before closure, \textsc{Step} updates $E$ to $E\cup\{a\}$, so
inclusion orders the append-only states. A mutable or sequence-valued record
would require a different state space.

\begin{definition}[Grounding lattice]
\label{def:grounding-lattice}
For a fixed task $\mathcal T$, let $\mathcal A_{\mathcal T}$ be the
set of normalized nonterminal atoms the task can store. The
\newterm{grounding lattice} is the powerset lattice
\[
  \mathcal{L}_{\mathcal T} =
  (\mathcal{P}(\mathcal A_{\mathcal T}),\subseteq),
\]
with union as join and $\bot_C=\emptyset$ as bottom
(\autoref{sec:background:lattices}). Every pre-close state lies in this
lattice, but admission controls which states a run can reach. The lattice
therefore specifies how stored evidence is ordered and combined, not which
proposals must be accepted or how the model searches. A state is
\newterm{complete} for $v$ when \textsc{Close} licenses $v$; completeness is
not the lattice top.
\end{definition}

\begin{wrapfigure}{R}{0.4\linewidth}
\centering
\resizebox{0.96\linewidth}{!}{%
\input{figures/lattice-cube}%
}
\caption{A toy grounding lattice for three binary slots.}
\label{fig:lattice-cube}
\end{wrapfigure}
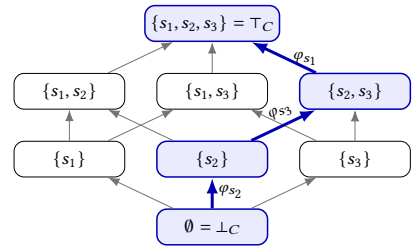

In \autoref{fig:lattice-cube}, three one-atom slots yield eight
subsets, and the blue edges show one construction order. Unioning the same
atoms in another order yields the same set. Real tasks may store multiple atoms
per slot and project them to the finite grade chain below, while the
certificate remains ordered by atom inclusion.

A task chooses a small vocabulary of records for this state space.
The base design distinguishes two nonterminal atom kinds and one terminal
record kind:
\begin{itemize}[noitemsep,topsep=2pt,leftmargin=*]
\item a \newterm{think atom} $\langle s,\mathit{body}\rangle$, which files a
      claim and its evidence in slot $s$;
\item an \newterm{expand atom} $\langle s,s'\rangle$, which opens a justified
      sub-slot $s'$ below $s$; admission permits it only when $s$ is
      an available slot, so an admitted expansion may add an available slot and
      extending $E$ cannot remove one; and
\item a \newterm{commit atom} $\langle v\rangle$, which
      \textsc{Close} adds only after licensing $v$ and which seals the
      certificate without contributing an evidence grade.
\end{itemize}
The harness returns refusal reasons to the model without changing
$E$. A task that permits execution may also expose a witness move: the model
proposes an entry point and input, and the harness runs it, records the
observation, and assigns the grade.

The shipped code uses the chain
\[
  \text{open}\;\sqsubseteq\;\text{filled}\;\sqsubseteq\;
  \text{cited}\;\sqsubseteq\;
  \text{witnessed}.
\]
An open slot has no grade-bearing record. A filled slot has model
prose but no recognized source reference. A cited slot has a
\texttt{file:line} reference that resolves in the evidence environment.
Licensing may separately bind that citation to another record, such as a
committed region. Neither check establishes that the prose follows from the
cited code. A witnessed slot contains an observation produced by harness
execution. The slot grade is the strongest grade assigned to any of its
records; lower-grade records remain in $E$.

The \newterm{grade projection} $\pi(E)$ of a certificate keeps only the
available slots, $S(E)$, which are the initial slots $S_0$ plus every
admitted expansion, and each slot's strongest grade. It is the progress view
shown to the model, and its number of strict increases is bounded
(\autoref{prop:bounded-ascent}). Licensing reads the full certificate,
including bodies, typed outcomes, resolutions, and task-derived obligations,
not only this projection.

\subsection{Admission, Transition, and Composition}
\label{sec:method:join}

The harness exposes the available proposal kinds as certificate
tools. The model chooses which tool to call, but only an admitted proposal
changes $E$.

\begin{definition}[Certificate tools]
\opThink proposes a body for a slot, \opExpand proposes a sub-slot
(a task with no expansion budget does not offer it), and
\opCommit proposes a verdict. Read-only file tools are observations within a
move and do not change the certificate. A task may additionally expose
\opWitness; only the harness may execute it and assign the resulting grade.
The move set a task exposes is these base tools plus any
task-specific proposal kind.
\end{definition}

Admission is a deterministic function of the evidence environment,
the current certificate, and the proposal. The harness refuses proposals that
violate task rules, such as targeting a nonexistent slot or contradicting
better-graded evidence, and returns a refusal reason.

\begin{figure}[t]
\centering
\input{figures/refused}%
\caption{The licensing check at the model--harness boundary. Each column
reads from certificate state $E$, through model proposal $p$, to the harness
decision. \textsc{Close} refuses a verdict that contradicts $E$ or leaves an
obligation open. Prose that claims reproduction remains cited because only a
successful harness-run witness can earn the witnessed grade.}
\Description{Three vertical trajectories show a contradicted commit, a commit
with an open obligation, and a model claim that cannot self-assign the
witnessed grade. Each trajectory ends in a harness refusal.}
\label{fig:refused}
\end{figure}
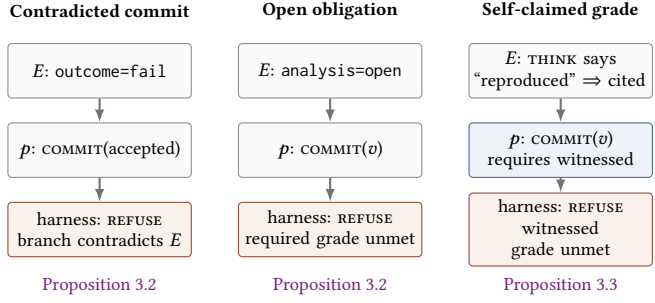

We follow the licensing check through three concrete cases in
\autoref{fig:refused}. In the first two, the commit proposal is well-formed in isolation, but
\textsc{Close} reads the accumulated certificate $E$. The rule refuses an accepted
verdict after a failing outcome and refuses any verdict while a required
analysis remains open. In the third, the model writes that it reproduced an
event, but prose can earn only the cited grade, so the witnessed obligation
remains open until the harness runs the witness itself.

The first two refusals are what a schema-only gate misses: the proposal's
shape is valid, but its relation to $E$ is not. The third marks the other half
of the trust boundary: the model may propose evidence, but only the harness
can award the grade attached to it. Grade authority is therefore the premise
on which witnessed soundness relies.

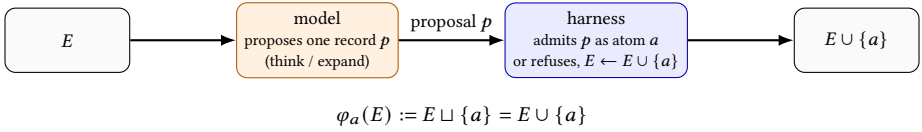
\begin{figure}[t]
\centering
\input{figures/agent-step}%
\caption{How the harness decides at runtime which proposed record to
admit and which to refuse. The model proposes a record through a
certificate tool; the harness admits it as atom $a$ and unions it into
$E$, or refuses it. The harness, not the model, performs the
union.
}
\label{fig:agent-step}
\end{figure}

\begin{definition}[Certificate join and accepted-record update]
\label{def:join}
The join of two pre-close atom sets is their union,
$E_1\sqcup E_2 := E_1\cup E_2$. For any nonterminal atom $a$, define
$\varphi_a(E):=E\cup\{a\}$. When admission accepts a proposal as $a$,
\textsc{Step} applies this update.
\end{definition}
Set union makes the join associative, commutative, and idempotent,
and each $\varphi_a$ extensive, monotone, and idempotent. These properties
describe the update, not admissibility: an atom accepted from one state need
not be accepted from every larger state, and some lattice elements may be
unreachable. An admitted think atom raises a slot's grade only if it passes a
stronger check; no update removes an earlier record. \autoref{fig:agent-step}
shows one proposal and admission step.

The set representation also yields a secondary rechecking
property.
\begin{proposition}[Fixed-certificate rechecking]
\label{prop:fixed-certificate-rechecking}
Fix a task and evidence environment, and suppose \textsc{Close}
licenses $v$ from the pre-close atom set $A$. Unioning the elements of $A$
into $\bot_C$ in any order reconstructs exactly $A$. Because grading,
resolution, obligations, and licensing are deterministic functions of $A$,
\textsc{Close} licenses $v$ again.
\end{proposition}

The claim follows from associativity and commutativity of union and
state-local determinism. It rechecks fixed certificate
contents; it does not rerun admission, tools, or the model's search, and a
different live order may produce a different admitted set.
We also audit that the shipped implementation honors this
property, since a bug that reads filing position would silently make the
printed verdicts depend on arrival order. From the evaluation runs we take
\Opus{} and \Qwen{} journals that span both verdicts, multiple regions,
and both acceptance and localization certificates. For each
journal we keep the recorded commit, erase the order of the pre-close atoms,
and rebuild the certificate under \num{25} seeded permutations. Every permutation yields the same grades, resolutions, obligations, and
licensing decision, confirming that the harness recomputes the same decision
from a fixed certificate without a model session. The audit covers the verdict, not the
trajectory: a resumed model need not generate the same continuation.
The binary join would also merge independently built certificates, but
every evaluated run constructs one certificate sequentially, so concurrent
certificate construction lies outside this paper's scope.

\subsection{Agent Loop as Worklist Iteration}
\label{sec:method:loop}

\begin{algorithm}[t]
  \caption{Budgeted certificate construction. The model chooses a
  proposal; the harness changes $E$ only after admission and returns an answer
  only after \textsc{Close} succeeds.}
  \label{alg:worklist}
  \input{algorithms/worklist}
\end{algorithm}

After each round, the harness gives the model the admitted
certificate, current grades, task-defined guidance over unresolved slots, and
any refusal from the preceding proposal. The model uses this information to
choose its next proposal and when to attempt a verdict. Once it proposes
$v$, $W(E,v)$ denotes its grade obligations that are not yet met. The
lattice orders the stored states; it does not schedule the model. Other
licensing checks may still refuse $v$ when $W(E,v)$ is empty.

The loop in \autoref{alg:worklist} starts from the blank
certificate. An admitted non-commit proposal adds one atom and never lowers
the grade projection, while a refusal leaves $E$ unchanged. The model chooses
when to call \opCommit; \textsc{Close} either refuses it or seals the
certificate. If the budget expires first, the run returns \verdictNone, scored
as wrong. Stable grades or empty guidance do not end the run.
The loop is worklist-like: unresolved requirements identify remaining
work, the model proposes records, and the harness unions admitted records into
the state. Unlike classical fixed-point worklist evaluation, it does not apply
a fixed family of transfer functions until no pending update changes the state.
The model may propose records outside the reported guidance, and the run ends
when \textsc{Close} licenses a verdict or the budget expires.

The correspondence with abstract interpretation is structural rather
than semantic, and \autoref{sec:method:limitations} states what it does
and does not carry over.

\subsection{Verdict Licensing}
\label{sec:method:verdict}

\textsc{Close} is the sole route from certificate to answer. The model
may propose $v$, but the harness records it only if
$\mathsf{lic}_{\mathcal{T}}(\Gamma,E,v)$ holds. For patch acceptance, the
predicate checks every required behavioral region and the branch their outcomes
imply; for fault localization, it binds each committed region to its own cited
analysis and enforces a span bound. Both instantiations read their obligations
off the record rather than off a preprocessing pass. Patch acceptance floors an
instance's region count at two when its own premise slots list no tests, and
fault localization owes one cited analysis for each region the agent admitted.
Because both derived facts
are conditions inside $\mathsf{req}_{\mathcal{T}}$ rather than atoms on file, a
third party recomputes either obligation set by running
$\mathsf{req}_{\mathcal{T}}$ over the certificate, and the grounding lattice
needs no closure-system refinement to derive them.
\autoref{fig:cert-template} shows one template per evaluated task. In
patch acceptance, each sampled test becomes a region with a cited analysis, an
existence field, and an outcome the gate reads; the accept branch, which is
cheap to assert, also owes an outcome audit and an account of untraced tests.
Fault localization drops the verdict branch and binds each rank to its own
cited analysis. Bug witnessing adds a \texttt{WITNESS} slot that only the
harness can write.

\begin{figure}[t]
\centering
\begin{subfigure}{\linewidth}
  \centering
  \input{figures/patch-cert-template}%
  \caption{Patch acceptance.}
  \label{fig:patch-cert-template}
\end{subfigure}

\begin{subfigure}[t]{0.495\linewidth}
  \centering
  \input{figures/faultloc-cert-template}%
  \caption{Fault localization.}
  \label{fig:faultloc-cert-template}
\end{subfigure}\hfill
\begin{subfigure}[t]{0.495\linewidth}
  \centering
  \input{figures/witness-cert-template}%
  \caption{Bug witnessing (crash and leak).}
  \label{fig:witness-cert-template}
\end{subfigure}
\caption{The certificate template of each evaluated task, condensed. Each
field is a slot or slot family; \texttt{Region.R\{n\}} and
\texttt{Claim.k} are repeated records whose children carry their own
obligations. The harness enforces each gate; task instructions supply each
field's natural-language meaning.}
\Description{Three monospaced templates. Patch acceptance lists definition,
premise, region, adversarial, and conclusion fields and a gate that maps region
outcomes to a verdict. Fault localization lists ranked regions with cited
analyses and a gate that binds each rank to its analysis. Bug witnessing lists
claims, a harness-only WITNESS slot, and a gate that requires a certified
witness.}
\label{fig:cert-template}
\end{figure}

Separating proposal from judgment makes the final check independent
of the model transcript. Given a fixed task, environment, and admitted atoms,
a third party can recompute grades, obligations, and licensing.
Two guarantees follow from this separation.

\begin{proposition}[Licensing obligation invariant]
\label{prop:grounding-safety}
If $\mathcal{T};\Gamma\vdash E\Downarrow v$, then every
$(s,g)\in\mathsf{req}_{\mathcal{T}}(E,v)$ has a recorded grade
$g'\sqsupseteq_G g$, and no settled outcome in $E$ contradicts $v$.
\end{proposition}

\begin{proposition}[Grade authority]
\label{lem:grade-authority}
If admitting a proposal raises slot $s$ to grade $g'$, then the resulting
atom passed the task-defined harness check required to assign $g'$.
\end{proposition}

Both hold by construction: the invariant is the contract imposed on
$\mathsf{lic}_{\mathcal{T}}$, and only the harness assigns grades. Their force
therefore depends on what the harness trusts: the deterministic reducer, the
task definition with its obligations and licensing rule, the source reader over
$\Gamma$, and, for witnessed claims, the executor and the task's observation
oracle. The model, its transcript, and every atom it proposes are untrusted
inputs, and a defect in a trusted component lies outside these guarantees.

\subsection{Limitations of the Correspondence}
\label{sec:method:limitations}

The construction above is structurally analogous to abstract
interpretation: certificate states are ordered, admitted-record updates
accumulate by union, and unmet obligations identify pending work. This is a structural analogy,
not the semantic soundness contract of abstract interpretation.

Unlike classical worklist iteration, \AgInt supplies neither a fixed
scheduler nor a mechanical fixed-point stopping test. A classical worklist
terminates when no pending update can change its state. In \AgInt, the model
chooses when to propose a verdict and the harness then checks it with
\textsc{Close}; a model policy may never propose a licensed finish, so finite
height does not imply successful completion
(\autoref{sec:method:loop}).
Like deployed agent systems
\citep{yang2024sweagent,zhang2024autocoderover,shinn2023reflexion}, \AgInt
therefore enforces resource caps. A licensed commit returns a verdict,
while finite operational budgets halt an unsuccessful run with no
verdict. The calculus proves only that the harness-visible grade
projection can strictly increase boundedly many times, as follows.

Raw atom inclusion has no finite height, because the model can file
arbitrarily many distinct bodies without changing a slot's grade. Refusals,
by contrast, leave $E$ unchanged.
Rather than force convergence on this raw domain, the harness bounds
the grade projection $\pi(E)$, which discards those
projection-inert additions. Admission refuses an expand proposal past the
bound $B_{\mathcal{T}}$, so every reachable state carries at most
$|S_0|+B_{\mathcal{T}}$ slots.

\begin{proposition}[Bounded ascent of the grade projection]
\label{prop:bounded-ascent}
Assume finite $S_0$, a finite grade chain, at most $B_{\mathcal{T}}$
admitted expansions, and a fixed task whose state is append-only and
whose grading is deterministic and monotone in $E$. Order
projections by $\pi(E)\sqsubseteq\pi(E')$ exactly when
$S(E)\subseteq S(E')$ and
$\mathsf{grade}(\Gamma,E,s)\sqsubseteq_G\mathsf{grade}(\Gamma,E',s)$ for
every $s\in S(E)$. If $k$ is the number of strict grade increases available
above open, any run induces an ascending projection sequence with at most
$k|S_0| + (k+1)B_{\mathcal{T}}$ strict increases.
\end{proposition}

The bound concerns progress in $\pi(E)$, not completion: a
projection-inert move may still append a body or change a licensing
query, so a licensed commit, never projection stability, ends a run. The
cited threshold gives $k=2$ and witnessed gives $k=3$, and the headline
pure-reasoning profiles run at $B_{\mathcal{T}}=0$, so their projections
make at most $k|S_0|$ strict increases.

The witnessed grade provides one-sided evidence: a
harness-reproduced fault confirms that the declared event occurred in that
execution, while a clean run establishes
only that the declared event did not occur in that execution, not that the
fault is globally absent. Below witnessed, \AgInt remains ``soundy'' rather
than sound in the abstract-interpretation sense
\citep{livshits2015soundiness}. Its guarantees concern the certificate and
harness, not the truth of every model interpretation.

%% file: figures/lattice-cube.tex
\begin{tikzpicture}[
  every node/.style={font=\small, inner sep=2pt},
  cov/.style={thin,-{Latex[length=1.6mm]}, draw=black!55},
  kleene/.style={very thick, -{Latex[length=2.2mm]}, draw=blue!70!black},
  cls/.style={draw, rounded corners, minimum height=0.5cm,
              minimum width=1.55cm, align=center, font=\scriptsize},
  hot/.style={draw=blue!70!black, line width=0.6pt, fill=blue!8,
              rounded corners, minimum height=0.5cm,
              minimum width=1.55cm, align=center, font=\scriptsize},
  x=1.00cm, y=0.95cm
]
  \node[hot] (b)   at (0,    0) {$\emptyset = \bot_C$};
  \node[cls] (s1)  at (-2, 1) {$\{s_1\}$};
  \node[hot] (s2)  at ( 0, 1) {$\{s_2\}$};
  \node[cls] (s3)  at ( 2, 1) {$\{s_3\}$};
  \node[cls] (s12) at (-2, 2) {$\{s_1,s_2\}$};
  \node[cls] (s13) at ( 0, 2) {$\{s_1,s_3\}$};
  \node[hot] (s23) at ( 2, 2) {$\{s_2,s_3\}$};
  \node[hot] (t)   at ( 0, 3) {$\{s_1,s_2,s_3\} = \top_C$};

  \draw[cov] (b) -- (s1);
  \draw[cov] (b) -- (s3);
  \draw[cov] (s1) -- (s12);
  \draw[cov] (s1) -- (s13);
  \draw[cov] (s2) -- (s12);
  \draw[cov] (s3) -- (s13);
  \draw[cov] (s3) -- (s23);
  \draw[cov] (s12) -- (t);
  \draw[cov] (s13) -- (t);

  \draw[kleene] (b) -- node[right, font=\tiny, pos=0.55]
    {$\varphi_{s_2}$} (s2);
  \draw[kleene] (s2) -- node[above, font=\tiny, pos=0.55, sloped]
    {$\varphi_{s_3}$} (s23);
  \draw[kleene] (s23) -- node[right, font=\tiny, pos=0.45]
    {$\varphi_{s_1}$} (t);
\end{tikzpicture}

%% file: figures/refused.tex
\begin{tikzpicture}[
  every node/.style={font=\scriptsize, inner sep=2.5pt},
  trajstep/.style={draw=black!45, rounded corners=1.5pt, align=center,
                   text width=2.25cm, minimum height=0.72cm, fill=black!2},
  refusal/.style={trajstep, draw=BrickRed!70!black, fill=BrickRed!5},
  authority/.style={trajstep, draw=NavyBlue!70!black, fill=NavyBlue!6},
  flow/.style={thick, -{Latex[length=1.7mm]}, draw=black!55}
]
  \node at (0, 0.55) {\textbf{Contradicted commit}};
  \node[trajstep] (a1) at (0, -0.25)
    {$E$: \texttt{outcome=fail}};
  \node[trajstep] (a2) at (0, -1.30)
    {$p$: \textsc{commit}(\vAccepted)};
  \node[refusal] (a3) at (0, -2.35)
    {harness: \textsc{refuse}\\branch contradicts $E$};
  \draw[flow] (a1) -- (a2);
  \draw[flow] (a2) -- (a3);
  \node at (0, -3.10) {\autoref{prop:grounding-safety}};

  \node at (3.05, 0.55) {\textbf{Open obligation}};
  \node[trajstep] (b1) at (3.05, -0.25)
    {$E$: \texttt{analysis=open}};
  \node[trajstep] (b2) at (3.05, -1.30)
    {$p$: \textsc{commit}($v$)};
  \node[refusal] (b3) at (3.05, -2.35)
    {harness: \textsc{refuse}\\required grade unmet};
  \draw[flow] (b1) -- (b2);
  \draw[flow] (b2) -- (b3);
  \node at (3.05, -3.10) {\autoref{prop:grounding-safety}};

  \node at (6.10, 0.55) {\textbf{Self-claimed grade}};
  \node[trajstep] (c1) at (6.10, -0.25)
    {$E$: \textsc{think} says\\``reproduced'' $\Rightarrow$ cited};
  \node[authority] (c2) at (6.10, -1.30)
    {$p$: \textsc{commit}($v$)\\requires witnessed};
  \node[refusal] (c3) at (6.10, -2.35)
    {harness: \textsc{refuse}\\witnessed grade unmet};
  \draw[flow] (c1) -- (c2);
  \draw[flow] (c2) -- (c3);
  \node at (6.10, -3.10) {\autoref{lem:grade-authority}};
\end{tikzpicture}

%% file: figures/agent-step.tex
\begin{tikzpicture}[
  every node/.style={font=\small, inner sep=3pt},
  box/.style={draw, rounded corners, align=center,
              minimum height=0.85cm, font=\scriptsize},
  agent/.style={box, fill=orange!12, draw=orange!70!black,
                minimum width=2.0cm},
  system/.style={box, fill=blue!8, draw=blue!70!black,
                 minimum width=2.0cm},
  state/.style={draw, rounded corners, align=center, font=\scriptsize,
                minimum height=0.85cm, minimum width=1.65cm,
                fill=black!2},
  flow/.style={thick, -{Latex[length=2mm]}},
  atom/.style={font=\scriptsize, align=center}
]
  \node[state] (in) at (0, 0) {$E$};
  \node[agent, right=1.4cm of in] (ag) {model\\\tiny proposes one record $p$\\\tiny (think / expand)};
  \node[system, right=1.4cm of ag] (sy) {harness\\\tiny admits $p$ as atom $a$\\\tiny or refuses, $E \gets E \cup \{a\}$};
  \node[state, right=1.4cm of sy] (out) {$E \cup \{a\}$};
  \draw[flow] (in) -- (ag);
  \draw[flow] (ag) -- node[atom, above]{proposal $p$} (sy);
  \draw[flow] (sy) -- (out);
  \node[font=\footnotesize] at ($(in.west)!0.5!(out.east) + (0,-1.0)$)
    {$\varphi_a(E) := E \sqcup \{a\} = E \cup \{a\}$};
\end{tikzpicture}

%% file: algorithms/worklist.tex
\begin{algorithmic}[1]
\State $E \gets \emptyset = \bot_C$
\Comment{the blank template}
\While{a budget remains}
  \State provide $E$, grades, and unresolved-slot guidance to the model
  \Comment{the harness informs; it does not schedule}
  \State the model proposes one move from the task's move set
  \Comment{reads observe and do not move the projection}
  \If{the move proposes a verdict $v$}
    \If{\textsc{Close} licenses $v$}
      \State persist $\mathsf{seal}(E,v)$
      \Comment{record the terminal commit}
      \State \Return $v$
    \Else
      \State report a closure refusal; $E$ is unchanged
    \EndIf
  \ElsIf{\textsc{Admit} accepts the proposal as atom $a$}
    \Comment{deterministic in evidence, state, and proposal}
    \State $E \gets E \sqcup \{a\}$
    \Comment{monotone: the grade projection rises or is unchanged}
  \Else
    \State report the refusal; $E$ is unchanged
    \Comment{no grade is ever lowered}
  \EndIf
\EndWhile
\State \Return \verdictNone
\Comment{budget exhausted before closure; scored as a wrong answer}
\end{algorithmic}

%% file: figures/patch-cert-template.tex
\begin{tcolorbox}[
  colback=white,
  colframe=black!60,
  width=\linewidth,
  boxrule=0.5pt,
  arc=2pt,
  left=2pt, right=2pt, top=1pt, bottom=1pt
]
\scriptsize
\begin{verbatim}
DEFINITIONS  D1  ACCEPTED iff every listed F2P test passes and every listed P2P
                 test stays passing (no listed tests: a concrete bug-to-fix trace)
PREMISES     P1  what the patch changes   P2  test-patch summary   P3/P4  F2P/P2P sets
REGIONS      one per sampled listed test (>= 2 when no tests are listed):
  Region.R{n}            the region and why it is reachable
  Region.R{n}.Analysis   patched vs. correct behavior -- MUST cite a [file:line]
  Region.R{n}.Existence  [PRESENT/ABSENT]; required for PASS -> ACCEPT
  Region.R{n}.Outcome    [PASS/FAIL] under D1
ADVERSARIAL  DevilsAdvocate  strongest case for the opposite verdict; names the
                             least-certain region
             OutcomeAudit    iff every Outcome = PASS: re-derive that region's outcome
             ResidueAccount  before ACCEPT: untraced tests + cited discharge, or zero
CONCLUSION   COUNTEREXAMPLE | NO COUNTEREXAMPLE EXISTS;  [ACCEPTED/NOT ACCEPTED]
GATE         any settled FAIL => REJECTED;  all PASS => ACCEPTED only if the audit
             passes, each PASS test exists, and the residue is discharged
\end{verbatim}
\end{tcolorbox}

%% file: figures/faultloc-cert-template.tex
\begin{tcolorbox}[
  colback=white,
  colframe=black!60,
  width=\linewidth,
  boxrule=0.5pt,
  arc=2pt,
  left=2pt, right=2pt, top=1pt, bottom=1pt
]
\scriptsize
\begin{verbatim}
DEFINITIONS
 D1 a region is correct iff it overlaps a
    developer-fix region in the same file
 D2 scope: failing test, loaded classes
PREMISES  P1 failing test: assertion, path
          P2 passing tests, or "none"
          P3 candidate files/classes
REGIONS  in descending suspicion:
 Region.R{n}           range + hypothesis
 Region.R{n}.Analysis  MUST cite [file:line]
CONCLUSION  why R1, R2, ... are so ordered
COMMIT      ranked [(file, start, end), ...]
GATE rank k binds to Region.Rk.Analysis:
       same file, overlapping citation
     merge same-file ranges <= 5 lines apart
     refuse any range wider than 40 lines
\end{verbatim}
\end{tcolorbox}

%% file: figures/witness-cert-template.tex
\begin{tcolorbox}[
  colback=white,
  colframe=black!60,
  width=\linewidth,
  boxrule=0.5pt,
  arc=2pt,
  left=2pt, right=2pt, top=1pt, bottom=1pt
]
\scriptsize
\begin{verbatim}
DEFINITIONS
 D1 WITNESSED iff a submitted program makes
    the harness observe the flagged fault
    (crash or leak) at the flagged file:line
 D2 scope: what a caller can supply via API
PREMISES  P1 flagged site   P3 entry points
          P2 bad value      P4 guards on path
          (leak: resource + its close duty)
CLAIMS  Claim.k: path entry -> site; guards
        fail; concrete input -- MUST cite
WITNESS   harness-written; model cannot fill
NO_COUNTEREXAMPLE  paths tried, what blocked
CONCLUSION  WITNESSED | NO WITNESS FOUND
GATE witnessed => WITNESS atom the oracle
       certified at site (leak: OS handle)
     no witness found => ungated
\end{verbatim}
\end{tcolorbox}

%% file: src/experiments.tex
\section{Experiments}
\label{sec:experiments}

Our evaluation answers three research questions:
\begin{enumerate*}[label=\textbf{RQ\arabic*},ref=RQ\arabic*]
\item \textbf{(effectiveness)} does the grounding calculus improve agentic
reasoning on program-centric analysis tasks?
\item \textbf{(mechanism)} how do the scaffold and the certificate body
divide \AgInt's gains?
\item \textbf{(witnessing)} with execution licensed, can the same calculus
reproduce faults from externally proposed warnings?
\end{enumerate*}
The first two questions use execution-free patch acceptance and fault
localization, which we score against recorded test outcomes and developer
fixes. The third enables execution and asks for concrete witnesses of
static-analyzer warnings.

\paragraph{Implementation}
Because the agent harness can affect results
\citep{yang2024sweagent,kapoor2025aiagents,kapoor2025hal}, we hold the SDK,
model endpoint, tool access, and budget fixed while varying how the harness
represents and licenses evidence. Every experiment runs on the Strands agent SDK
\citep{liguori2025strands}, following the practice of
\citet{yang2024sweagent}, and every model is an Amazon Bedrock endpoint. For
the two execution-free tasks, the agent receives \opThink{} and \opCommit{}
without \opExpand{}, because the task surface is narrow enough without it,
and it can read a read-only
checkout but not change it (\texttt{read\_file}, \texttt{grep}, and
\texttt{glob}). Their templates already repeat a region record as often as an
instance needs. Granting \opExpand{} back gains nothing beyond the noise band
on any of the four datasets with \Opus{} or \GLM{}, yet costs \num{1} to
\num{10} percent more steps per record. For bug witnessing, the agent additionally receives \opExpand{},
\opWitness{}, and sandboxed code execution, so it can collect evidence for a
proposed bug and then submit the entry point and input on which the harness
reproduces it. Both configurations implement the calculus of
\autoref{sec:method}, which the tables write as \AgInt{}.

\subsection{Evaluation Setting}
\label{sec:experiments:setting}

\paragraph{Compared methods}
On the two execution-free tasks, three methods share the model, the harness
above, and a \num{60}-turn budget. The \Bare{} baseline uses the task
prompt alone. The \Acr{} baseline fills its task-specific certificate as prose
and commits without a gate \citep{ugare2026acr}; \AgInt{} uses typed records,
graded admission, and the \opCommit{} licensing gate. Thus the arms differ in
how they manage evidence, not in model or tool access. We test the open-weight
\Qwen{}, \GLM{}, and \Minimax{}, plus \Opus{} as a capability ceiling.
For the real-world bug-witnessing comparison, both \AgInt{} and the matched
shell-agent baseline use \GLM{}. The shell agent runs on the same Strands
harness but replaces the certificate and the graded witness tool with a plain
shell, under the same sandbox, tree, classpath, budget, and oracle.

\subsubsection{Patch Acceptance}
\label{sec:setting:patch-acceptance}

\emph{Given a patch and unit-test suite, will the patch pass?} Running the
suite settles the question, but each verdict can require a checkout, build,
container, and test run. Training loops evaluate candidate patches at
$10^5$ scale per run and CI repeats the cost across commits
\citep{datacurve2026deepswe,jain2025r2egym,pan2024swegym}, motivating an
execution-free judgment.

\paragraph{Task definition}
An instance follows the SWE-bench convention~\citep{jimenez2024swebench}:
a repository at the pre-fix commit, a patch, and the fail-to-pass and
pass-to-pass test sets. Each instance carries a gold label recorded by
executing the suite before any agent runs, \vAccepted{} when every test in both
sets passed on the patched program and \vRejected{} otherwise. The agent
receives the repository, the patch as a unified diff, and the test bodies as
static artifacts, and it returns the verdict without invoking any interpreter,
compiler, or test runner. The task therefore instantiates
$V_{\mathcal{T}} = \{\vAccepted, \vRejected\}$ with the cited grade as its
threshold and no execution move. The agent sees the test bodies but not their
outcomes, and never the developer's gold patch, so it sees what a training loop
or CI gate would see.

\paragraph{Datasets}
Both splits come from agentic-coding benchmarks whose repositories
postdate the models under test, because a judge that has seen the fixing commit
can recognize the answer instead of deriving it. Every patch carries the label
its recorded containerized run produced, the contract an RL reward or a CI gate
applies. \SBL~\citep{li2026repolaunch} supplies issue-level tasks with
community-agent patches across multiple languages. We keep the \num{108}
instances whose community submissions disagree, because a mixed-verdict
instance forces a read of the candidate patch rather than an instance-level
difficulty prior, and we add a fixed seed-1 sample of \num{20} rejected patches
from never-solved instances to remove the accept skew ($n=\num{128}$, 64/64).
\DSW~\citep{datacurve2026deepswe} supplies the harder split. Its \num{113}
tasks were authored against live repositories in 2026 and canary-marked against
training use, its labels are per-rollout verifier output, and its median
solution spans \num{844} lines. Because one configuration runs up to four times
on a task and the rollouts often land on opposite labels, we pair an accepted
with a rejected rollout of the same configuration on the same task
($n=\num{200}$, 100/100 over \num{73} tasks). Each pair is near-identical in
surface form but divergent in behavior, so the majority-class baseline is
\num{0.500}. \DSW{} tasks are mostly feature requests whose written
specification is the acceptance criterion, so the agent also sees that
specification.

\paragraph{Metrics}
Following the convention
for asymmetric-cost detection~\citep{he2009imbalanced,menzies2007defectprediction},
we treat rejecting a bad patch as the positive event. LLM judges approve far
more readily than they flag
errors~\citep{sharma2024sycophancy,thakur2024judgingjudges,kamoi2024llmerrordetection},
and a false \textsc{accept} creates a hackable reward channel, whereas a false
\textsc{reject} only discards
samples~\citep{skalse2022rewardhacking,stroebl2024inferencescaling}. We
therefore lead with \FoneReject, the F1 score with \vRejected{} as the positive
class, and add the class-balanced \MacroFone{} and the \Accuracy on which prior
patch-acceptance work is scored. For instance $i$ with gold label $y_i$, let
$\hat{y}_i$ be the committed verdict. A run that commits neither label counts
as wrong, so we set $\hat{y}_i$ to the label opposite $y_i$ and every instance
enters the denominator. Counting true positives $\mathit{TP}$, false positives
$\mathit{FP}$, and false negatives $\mathit{FN}$ with \vRejected{} as the
positive class over the $n$ instances,
\begin{equation}
  \text{precision} = \frac{\mathit{TP}}{\mathit{TP}+\mathit{FP}},\qquad
  \RecallReject = \frac{\mathit{TP}}{\mathit{TP}+\mathit{FN}},\qquad
  \FoneReject = \frac{2\cdot\text{precision}\cdot\RecallReject}
                     {\text{precision}+\RecallReject},
  \label{eq:f1-reject}
\end{equation}
\begin{equation}
  \MacroFone = \tfrac{1}{2}\bigl(\FoneReject + \FoneAccept\bigr),\qquad
  \Accuracy = \frac{1}{n}\sum_{i=1}^{n}\mathbf{1}[\hat{y}_i = y_i],
  \label{eq:macro-accuracy}
\end{equation}
where \FoneAccept{} is \autoref{eq:f1-reject} with \vAccepted{} as the positive
class. Unqualified \FoneReject{} and \RecallReject{} name the reject class
throughout.

\subsubsection{Fault Localization}
\label{sec:setting:fault-localization}

\emph{Given a failing test, which source lines are responsible?} Execution has
established the symptom, but not the source location that must change, the
``last mile'' of debugging \citep{wong2016survey,chang2026lastmile}. Unlike
patch acceptance, localization also requires repository search.

\paragraph{Task definition}
An instance pairs a buggy program with a failing test that observes the
bug and a ground-truth fix location $R^\star$, the set of source line ranges
that the developer's fix modifies. The agent sees the source tree and the body
of the failing test, but no stack trace, error message, or other
execution-derived signal. It returns a ranked list $\hat{R}$ of predicted
\texttt{(file, start, end)} regions, most suspicious first, and
$\mathsf{lic}_{\mathcal{T}}$ binds each region to a cited analysis.
The gate also merges same-file regions within five lines and refuses
any region wider than \num{40} lines, above every gold region in the corpus,
so overlap cannot be bought with width (\autoref{fig:faultloc-cert-template}). A
prediction $r$ \newterm{covers} a gold region $g \in R^\star$ when the two name
the same file and their line ranges overlap. Overlap rather than exact match
keeps a deletion-only hunk, whose fixed line vanishes, coverable.

\paragraph{Datasets}
Two corpora of real Java bugs carry this task. For each bug we extract the
buggy revision, its failing test, and the gold location $R^\star$, the
line-level diff between the buggy and fixed revisions restricted to source
lines (no test files or comment-only changes), so ground truth comes from the
developer's fix at no extra execution cost. \Defects~v2.0.0
\citep{just2014defects4j} curates isolated, reproducible bugs from mature
open-source projects. We reproduce the sample of \citet{ugare2026acr} at random
seed~0 and report its \SplitLarge{} split ($n=\num{100}$); because the authors
publish no bug IDs, our sample is protocol-aligned rather than
instance-identical. We use the recorded triggering test and hold out tests that
the developer added solely to demonstrate the fix.
\GitBug~\citep{silva2024gitbugjava} collects \num{199} reproducible bugs from
Java commits made in 2023. We materialize the complete buggy repository rather
than only the classes the test loads, so \GitBug adds recent code, which bounds
what memorization of an older corpus could explain, and a whole-repository
search surface.

\paragraph{Metrics}
Following
AutoFL~\citep{kang2024autofl}, we apply reciprocal rank to line regions. For
instance $i$ with ranked prediction $\hat{R}^{(i)}$ and gold regions
$R^{\star}_i$, let $q_i$ be the first rank whose region covers some
$g \in R^{\star}_i$. Over $n$ instances,
\begin{equation}
  \MRR = \frac{1}{n}\sum_{i=1}^{n}\frac{1}{q_i},
  \label{eq:mrr}
\end{equation}
where $1/q_i = 0$ when no ranked region covers a gold region, so \MRR{}
reports how early the first true region surfaces. For inspection cost,
we adapt the ranked code-examination model of \citet{wong2016survey} to
bounded regions and count the lines a developer would read for nothing.
Writing $\hat{R}_N$ for the first $N$ ranks, $\lines(r)$ for the
\texttt{(file, line)} identities a region spans,
$\hat{P}_N = \bigcup_{r \in \hat{R}_N}\lines(r)$, and
$G = \bigcup_{g \in R^\star}\lines(g)$ for the gold lines, the wasted
inspection effort of one instance is
\begin{equation}
  \WastedAtN = |\hat{P}_N \setminus G|,
  \label{eq:wasted}
\end{equation}
which we average over instances; a run without a verdict inspects no
lines.
Prior work scores Top-$N$ region-overlap coverage
\citep{just2014defects4j,ugare2026acr}, which we also report:
\begin{equation}
  \AllAtN
  \;\stackrel{\Delta}{=}\;
  \mathbf{1}\bigl[\,\text{every } g \in R^\star \text{ is covered by
    some } r \in \hat{R}_N\,\bigr],
  \label{eq:faultloc-allat}
\end{equation}
averaged over instances. Because the hit test
asks only that two intervals intersect, a wide enough bracket scores
$\AllAt{5} = 1$ without localizing anything. We therefore report coverage at
rank one, where the top region must cover the fault by itself, and wasted
effort at rank five, which charges for the padding that a rank-one metric
never scores.

\subsubsection{Bug Witnessing}
\label{sec:setting:bug-witnessing}

\emph{Given a bug candidate, can the agent produce an input on which the
harness reproduces the suspected fault?} The task licenses execution and
reaches the \emph{witnessed} grade of \autoref{sec:method:lattice}. We
instantiate it in the analyzer-guided regime, where a static analyzer supplies
a \newterm{bug candidate}, a site that may contain a bug but has not been
confirmed.

\paragraph{Task definition}
A \newterm{witness} is a small executable reproducer that the agent
submits through \opWitness. The harness builds and runs it in a sandbox and
returns \emph{witnessed} only when a task-specific oracle confirms the claimed
fault at the claimed site, so an accepted witness acts as a one-sided oracle on
the candidate. \AgInt{} therefore returns \emph{witnessed} only for a candidate
whose claimed fault the harness has reproduced. When no witness emerges within
budget, it returns \emph{no-witness-found-in-budget}, an explicit cannot-decide
verdict, and never declares the candidate a false positive.
A run may submit up to \num{12} witnesses within a \num{200}-turn
budget (\autoref{fig:witness-cert-template}).

\paragraph{Observation oracles}
Two task-declared oracles decide the witnessed transition, one per warning
family, and neither reads a model-supplied field. For \emph{crashes}, the
harness attributes a thrown exception to the flagged site only when that site
is the deepest application frame on the causal chain, matched on file, line,
and the package declared in the flagged file. This rule refuses an exception
born in the runtime above the site, which a top-frame test would miss, and an
exception that first broke in project code below the site, which an
anywhere-on-stack test would over-attribute. Three forgery checks back it: the
trace must reach the default handler under the program's own entry frame, the
process must exit nonzero, and no frame may name a line past its file's end.
A \emph{resource leak} throws nothing, so the harness runs the witness under a
descriptor-tracking Java agent and a deterministic collector and inspects the
process when \texttt{main} returns. It certifies a leak only when an
operating-system file handle allocated at the flagged site is still open and is
not held by the test harness itself. A weaker tier flags an object whose
\texttt{close} never ran but that owns no operating-system handle; the two
tiers are reported apart, and the gate refuses a \emph{witnessed} commit backed
only by the weaker tier. Both oracles are one-sided in the must direction of
\autoref{sec:background:lattices}: a certified run confirms the fault at the
site, and a clean run licenses nothing. The guarantee concerns the observation,
not the entry surface, so the harness records each witness's entry visibility
beside the verdict rather than folding it into licensing.

\paragraph{Datasets and metric}
We run the task on two Java project pools, kept separate because their
composition drives leak witnessability: \num{20} projects from the
\num{132}-repository corpus of CodeCureAgent~\citep{joos2025codecureagent}, and
the \num{10} Apache Commons libraries. A project qualifies when it is
unarchived, recently pushed, redistributably licensed, and carries enough
main-tree warnings of the class; test-tree warnings are excluded. SonarQube
supplies null dereferences (\texttt{java:S2259}) and resource leaks
(\texttt{java:S2095}), and Infer supplies the matched issue types
\texttt{NULLPTR\_\allowbreak DEREFERENCE} and
\texttt{PULSE\_\allowbreak RESOURCE\_\allowbreak LEAK}. The two analyzers flag
largely disjoint sites even for matched classes, so we report their yields
separately. Over a fixed list $C$ of deduplicated candidates, we report
\begin{equation}
  \text{yield} = \frac{|\{c \in C \mid \text{the harness accepts a witness for } c\}|}{|C|}.
  \label{eq:witness-yield}
\end{equation}
Because the oracle is one-sided, yield lower-bounds the fraction of concrete
bugs among the candidates and does not estimate the analyzer's false-positive
rate.

\subsection{Patch-Acceptance Results}
\label{sec:experiments:reasoning}
\label{sec:experiments:acceptance}

The lead task asks whether grounding improves the prediction of a held-out test
outcome. We test it over both splits of \autoref{sec:setting:patch-acceptance}
and all four models. Two choices govern \AgInt's rows. The harness evicts raw tool
results once the model has filed the evidence it wants as atoms, and a
region whose outcome is \vPass{} must state whether the listed test exists
in the patched tree before it can license \vAccepted{}. \DSW{} adds one
requirement \SBL{} does not, an audit of the exact-equality assertion most
at risk, which leaves \SBL{} untouched and costs \Qwen{} about one standard
deviation of \FoneReject{}. We defer the matched-template control to
\autoref{sec:experiments:control}.

\begin{table}
  \caption{%
  Patch acceptance on the two execution-labelled datasets.
  Bold marks the best values.
  }
  \label{tab:patch-acceptance}
  \small
  \setlength{\tabcolsep}{2.5pt}
\input{tables/patch-acceptance}\end{table}

\AgInt{} leads both baselines on every metric, model, and dataset in
\autoref{tab:patch-acceptance}. The ordering survives \DSW{}, which weakens the
authorship cue that \SBL{} cannot exclude. Because both baselines accept nearly
everything, the gain comes from rejecting bad patches: \GLM's reject-class
recall nearly triples at the cost of a few false rejections. Capping \AgInt{}
at the plain agent's realized turn count leaves its score unchanged, so extra
steps do not explain the gain, although full \AgInt{} runs cost more per
instance. On \GLM{} and \Minimax{} the calculus sharpens completed judgments;
on \Qwen{} it mainly restores verdict production. At the \Opus{} ceiling the
margins are smallest and we claim no effect: on \SBL{}, all three methods fall
within a two-point \MacroFone{} window. The larger \DSW{} movement partly
reflects its benchmark-informed exact-equality mandate, which requires
confirmation on a new held-out surface.

\subsection{Fault-Localization Results}
\label{sec:experiments:faultloc}

The patch task answers a runtime question with a label. Fault
localization answers it with a location, so it tests whether grounding
buys precision and not just breadth. We run the
three methods on the two Java corpora of
\autoref{sec:setting:fault-localization} with all four models.
\label{sec:experiments:faultloc:surface}
The harness refuses to start a whole-repository instance until the
agent reads the checkout, a requirement a tool-log audit of an earlier
round motivated. It improves every arm and improves \AgInt{} most, as a
grounding discipline should when the evidence it consumes is restored.

\begin{table}
  \caption{Fault localization on the two datasets.
  Bold marks the best values.
  }
  \label{tab:faultloc}
  \small
  \setlength{\tabcolsep}{3pt}
\input{tables/faultloc}  \par\smallskip
  \raggedright\footnotesize
  $^{\ast}$\WastedAt{5} counts zero lines for a record without a
  verdict, so the \Qwen{} baselines' low values reflect their commit floor
  rather than tighter regions, and we bold no cell in that group.
\end{table}

\AgInt{} ranks the responsible lines above both baselines for every model and
corpus in \autoref{tab:faultloc}, except for a tie with the plain agent at the
\Opus{} ceiling on \GitBug{}; it nearly triples \GLM's rank-one hit rate on
\Defects{}. Its most consistent edge is inspection cost. Under \GLM{}, it cuts
the plain agent's wasted lines to about a third on \GitBug{}; under \Opus{}, to
about a quarter on both corpora despite thinner ranking margins. Under
\Minimax{}, the ranking and tightness leads persist while strict rank-one
coverage is level with the plain agent. The \Qwen{} baselines remain at a
commit floor, while \AgInt{} answers about four instances in five and every
repetition clears both baselines on each ranking metric. Relative to the
\Opus{} plain agent, this recovers about two fifths of the capability gap on
\Defects{} and one third on \GitBug{}. The region-width cap contributes to the
inspection-cost result.

\begin{takeaway}{RQ1 (effectiveness)}
The calculus improves both tasks below the model ceiling, where
\AgInt{} ranks first on reject-class F1 and \MRR{} and cuts bounded
inspection cost wherever the baselines commit. At the \Opus{} ceiling, the
patch-acceptance and ranking margins thin while the inspection-cost gain
persists, so the trend across model capability remains a hypothesis.
\end{takeaway}

\FloatBarrier

\subsection{Mechanism Ablation}
\label{sec:experiments:mechanisms}
\label{sec:experiments:control}

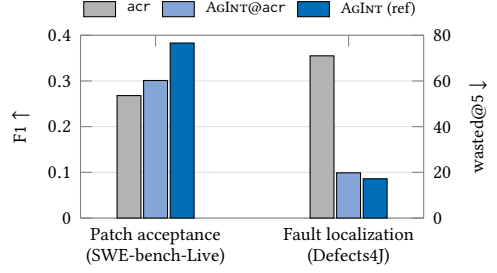
\begin{wrapfigure}[14]{r}{0.44\linewidth}
  \centering
  \input{figures/scaffold-body}%
  \caption{Scaffold-versus-body decomposition on \GLM.}
  \label{fig:scaffold-body}
\end{wrapfigure}
\AgInt{} differs from \Acr{} in both its agent loop and its certificate
design, so the main comparison cannot determine which change matters. To
separate them, we run \AcrBody{}, which keeps \Acr{}'s byte-identical
certificate body and gate-free commit rule but adds \AgInt{}'s persistent,
obligation-guided loop. We compare it with \AgIntRef{}, the reference
configuration for this ablation, which omits tool-result eviction and
accept-side obligations. The comparison from \Acr{} to \AcrBody{} isolates
the change to the loop, while the comparison from \AcrBody{} to
\AgIntRef{} isolates the change to the certificate and commit gate. We run
this comparison with \GLM{} on \SBL{} and the \Defects{} \SplitLarge{}
split (\autoref{fig:scaffold-body}).

On \SBL{}, F1 rises from \num{0.268} with \Acr{} to
\num{0.301} \pmsd{0.048} with \AcrBody{}, and then to \num{0.383} with
\AgIntRef{}. On \Defects{}, adding the persistent loop reduces
\WastedAt{5} from \num{71.0} to \num{19.8}, close to \num{17.2} with
\AgIntRef{}, and raises \MRR{} from \num{0.240} to \num{0.600}, compared
with \num{0.665}. The persistent loop therefore nearly reproduces
\AgIntRef{}'s fault-localization result, but not its patch-acceptance
result. The loop matters more for localization, while the certificate
and commit gate matter more for patch acceptance.

\begin{takeaway}{RQ2 (mechanism)}
The gains divide by task. The scaffold carries fault localization, and
the redesigned certificate body with its gate carries the reject-side
calibration on patch acceptance. Neither part explains both tasks.
\end{takeaway}

\FloatBarrier

\subsection{Bug-Witnessing Results}
\label{sec:experiments:witness}

\begin{table}
  \centering
  \caption{Bug-witness yield and efficiency using \GLM{} for both \AgInt{} and
  the matched shell-agent baseline on the fixed \num{30}-project corpus. \# is
  the pooled, deduplicated candidate count; rates are witnesses/\#. Efficiency
  values are averages over the candidates in their respective evaluations.}
  \label{tab:witness}
  \small
  \input{tables/witness}
\end{table}
Every positive carries a harness-reproduced observation under
the oracles of \autoref{sec:setting:bug-witnessing}. We use
SonarQube~\citep{sonarqube} and Infer~\citep{calcagno2015infer} warnings from
the fixed \num{30}-project corpus as null-dereference and resource-leak
candidates. SonarQube matches syntactic and dataflow patterns, while Infer
tracks nullness and resource ownership across procedure boundaries. The six
lists span both analyzers and bug classes, with null-dereference candidates
drawn from both project pools. Because each analyzer defines its own candidate
set, yields are compared within rows rather than across analyzers.

\paragraph{Witness yield}
\AgInt{} witnesses a clear majority of the pooled null-dereference sites,
and the harness rejects most remaining witness attempts without establishing
that the warnings are false under the one-sided oracle. Misses
are densest in the Commons SonarQube list and are reachability failures caught
by the oracle, not warnings the agent declined.

\paragraph{Baseline comparison}
To separate \AgInt's machinery from execution access, we compare it with
the shell agent of \autoref{sec:experiments:setting}, which runs on the
same Strands harness under the same sandbox, tree, classpath,
budget, and oracle. Pooled over the lists, \AgInt{} witnesses \num{293} of
\num{475} null-dereference candidates versus \num{253}, and \num{62} of
\num{117} resource-leak candidates versus \num{58}; the leak difference
remains small at this sample size (\autoref{tab:witness}). On the corresponding
evaluations, \AgInt{} uses \num{15.9} versus \num{24.5} executions per
candidate and costs \$\num{3.66} versus \$\num{8.83} per candidate. The yield
is higher on null-dereference candidates, while resource-leak yields are
similar at this sample size. The reported efficiency measurements likewise
favor \AgInt{}, with fewer executions and lower cost per candidate.

\begin{wraptable}{r}{0.42\linewidth}
  \centering
  \caption{Soundness-floor confusion matrix on the \num{142} matched
  Juliet pairs (\num{142} sites per row).}
  \label{tab:witness-soundness-floor}
  \small
  \input{tables/witness-soundness-floor}
\end{wraptable}
\paragraph{Soundness floor on a labelled corpus}
\label{sec:experiments:witness:soundness-floor}
The real-world warning lists lack labels, so no cell of
\autoref{tab:witness} tests whether the witnessed grade rejects a site that is
not a bug. The Juliet CWE-476/690 suite~\citep{boland2012juliet} supplies that
ground truth, because every case pairs a faulty method with a patched sibling.
We stage each side as a standalone project, strip every suite marker so the
agent cannot recognize the benchmark and answer from its label, and keep the
\num{142} pairs (\num{284} instances) whose two sides both compile in the
witness sandbox and whose patched side no caller can fault through public
writable state. We give each instance to \GLM{} once under the same
analyzer-guided protocol, crash oracle, and \num{12}-candidate budget.
\AgInt{} witnesses every faulty site and no patched site
(\autoref{tab:witness-soundness-floor}). The zero on the patched row records
refusal rather than abstention: the agent still submits candidates on \num{138}
of the \num{142} patched instances, and the harness refutes every one.
Hand-written forgery probes scope the claim. The crash oracle by itself would
certify two trace-shaped fabrications, and no run in any arm produced one, so
the zero false positives describe how the agent behaved under the harness
rather than what the harness precludes.

\begin{takeaway}{RQ3 (witnessing)}
With execution licensed, \AgInt{} witnesses \num{293} of \num{475} pooled
null-dereference candidates (\num{61.7} percent) and \num{62} of \num{117}
resource-leak candidates. A shell agent on the same harness trails by
\num{8.4} percentage points on null dereferences and matches the leak rate
within noise. Across the reported efficiency measurements, \AgInt{} uses
\num{0.65} times as many executions and reduces cost from \$\num{8.83} to
\$\num{3.66} per candidate. Each positive leaves a recheckable certificate.
The witnessed grade certifies the reproduced observation at the claimed site;
negative verdicts and real-caller reachability remain outside it.
\end{takeaway}

\subsection{Threats to Validity}
\label{sec:experiments:threats}

\paragraph{Stochastic baselines}
We repeat each \AgInt{} cell three times, but the \Bare{} and \Acr{}
cells are single runs because of time and budget limits. Because their
variation is unmeasured, we cannot place confidence intervals on
between-method differences, so we read the tables as descriptive rankings,
not as claims of statistical separation. The witnessing comparison likewise
rests on single draws per arm.

\paragraph{Sampling and scope}
The \DSW{} split contains \num{200} rollouts from \num{73} underlying
tasks, so task-level dependence makes the effective sample smaller than the
row count, and we report aggregate metrics rather than task-clustered
uncertainty. The corpora cover patch acceptance in recent multi-language
repositories and fault localization and bug witnessing in Java; they do not
establish a general law over tasks or model capability.

\paragraph{Guarantee boundary}
The two execution-free tasks stop at the cited grade, which checks that a
reference resolves but not that the cited code entails the model's claim. The
witnessed grade is narrow: the harness reproduces the claimed fault at the
claimed site but does not certify that a
public caller reaches it, and its one-sided oracle supports witness yield, not
recall or a false-positive rate. The controlled experiments therefore measure
the utility of a checked evidence discipline, not semantic soundness, and the
witnessing study measures reproduced defect yield, not a capability ranking.

%% file: tables/patch-acceptance.tex
  \begin{adjustbox}{max width=\columnwidth}
  \begin{tabular}{ll r@{}l r@{}l r@{}l r@{}l r@{}l r@{}l}
    \toprule
    & & \multicolumn{6}{c}{\SBL}
        & \multicolumn{6}{c}{\DSW} \\
    \cmidrule(lr){3-8}\cmidrule(lr){9-14}
    Model & Method
      & \multicolumn{2}{c}{\textbf{\FoneReject}}
      & \multicolumn{2}{c}{\textbf{\MacroFone}}
      & \multicolumn{2}{c}{\textbf{\Accuracy}}
      & \multicolumn{2}{c}{\textbf{\FoneReject}}
      & \multicolumn{2}{c}{\textbf{\MacroFone}}
      & \multicolumn{2}{c}{\textbf{\Accuracy}} \\
    \midrule
    \multirow{3}{*}{\Opus}
      & \Bare & 0.598 && 0.676 && 0.695 &
        & 0.537 && 0.631 && 0.655 & \\
      & \Acr  & 0.583 && 0.667 && 0.688 &
        & 0.581 && 0.668 && 0.690 & \\
      & \AgInt (ours)
        & \textbf{0.654} & \pmsd{0.011}
        & \textbf{0.701} & \pmsd{0.006}
        & \textbf{0.708} & \pmsd{0.005}
        & \textbf{0.619} & \pmsd{0.030}
        & \textbf{0.692} & \pmsd{0.018}
        & \textbf{0.710} & \pmsd{0.015} \\
    \cmidrule(lr){1-14}
    \multirow{3}{*}{\GLM}
      & \Bare & 0.256 && 0.465 && 0.547 &
        & 0.104 && 0.371 && 0.485 & \\
      & \Acr & 0.268 && 0.462 && 0.531 &
        & 0.056 && 0.356 && 0.495 & \\
      & \AgInt (ours)
        & \textbf{0.543} & \pmsd{0.068}
        & \textbf{0.616} & \pmsd{0.046}
        & \textbf{0.630} & \pmsd{0.040}
        & \textbf{0.173} & \pmsd{0.034}
        & \textbf{0.411} & \pmsd{0.024}
        & \textbf{0.507} & \pmsd{0.019} \\
    \cmidrule(lr){1-14}
    \multirow{3}{*}{\Minimax}
      & \Bare & 0.171 && 0.430 && 0.547 &
        & 0.019 && 0.330 && 0.475 & \\
      & \Acr & 0.162 && 0.411 && 0.516 &
        & 0.000 && 0.324 && 0.480 & \\
      & \AgInt (ours)
        & \textbf{0.494} & \pmsd{0.058}
        & \textbf{0.588} & \pmsd{0.031}
        & \textbf{0.609} & \pmsd{0.021}
        & \textbf{0.073} & \pmsd{0.002}
        & \textbf{0.362} & \pmsd{0.009}
        & \textbf{0.493} & \pmsd{0.018} \\
    \cmidrule(lr){1-14}
    \multirow{3}{*}{\Qwen}
      & \Bare & 0.032 && 0.062 && 0.062 &
        & 0.000 && 0.024 && 0.025 & \\
      & \Acr & 0.016 && 0.046 && 0.047 &
        & 0.010 && 0.034 && 0.035 & \\
      & \AgInt (ours)
        & \textbf{0.496} & \pmsd{0.040}
        & \textbf{0.416} & \pmsd{0.053}
        & \textbf{0.427} & \pmsd{0.051}
        & \textbf{0.467} & \pmsd{0.024}
        & \textbf{0.257} & \pmsd{0.016}
        & \textbf{0.317} & \pmsd{0.021} \\
    \bottomrule
  \end{tabular}
  \end{adjustbox}

%% file: tables/faultloc.tex
  \begin{adjustbox}{max width=\columnwidth}
  \begin{tabular}{ll rr r@{}l rr r@{}l rr r@{}l}
    \toprule
    & & \multicolumn{4}{c}{\AllAt{1} $\uparrow$}
      & \multicolumn{4}{c}{\MRR $\uparrow$}
      & \multicolumn{4}{c}{\WastedAt{5} $\downarrow$} \\
    \cmidrule(lr){3-6} \cmidrule(lr){7-10} \cmidrule(lr){11-14}
    Model & Corpus & \Bare & \Acr & \multicolumn{2}{c}{\AgInt}
      & \Bare & \Acr & \multicolumn{2}{c}{\AgInt}
      & \Bare & \Acr & \multicolumn{2}{c}{\AgInt} \\
    \midrule
    \multirow{2}{*}{\Opus}
      & \Defects & 0.370 & 0.390
          & \textbf{0.477} & \pmsd{0.021}
          & 0.711 & 0.805
          & \textbf{0.838} & \pmsd{0.010}
          & 69.0 & 41.0
          & \textbf{17.6} & \pmsd{0.9} \\
      & \GitBug & 0.231 & 0.206
          & \textbf{0.286} & \pmsd{0.025}
          & 0.820 & 0.804
          & 0.821 & \pmsd{0.014}
          & 48.6 & 33.1
          & \textbf{13.4} & \pmsd{0.7} \\
    \midrule
    \multirow{2}{*}{\GLM}
      & \Defects & 0.140 & 0.100
          & \textbf{0.400} & \pmsd{0.010}
          & 0.323 & 0.240
          & \textbf{0.633} & \pmsd{0.036}
          & 19.0 & 71.0
          & \textbf{17.7} & \pmsd{1.2} \\
      & \GitBug & 0.151 & 0.090
          & \textbf{0.221} & \pmsd{0.027}
          & 0.409 & 0.345
          & \textbf{0.618} & \pmsd{0.039}
          & 31.5 & 17.9
          & \textbf{11.7} & \pmsd{0.5} \\
    \midrule
    \multirow{2}{*}{\Minimax}
      & \Defects & 0.340 & 0.270
          & 0.327 & \pmsd{0.031}
          & 0.485 & 0.420
          & \textbf{0.535} & \pmsd{0.048}
          & 20.7 & 17.6
          & \textbf{15.9} & \pmsd{0.8} \\
      & \GitBug & 0.191 & 0.171
          & 0.194 & \pmsd{0.006}
          & 0.497 & 0.560
          & \textbf{0.598} & \pmsd{0.008}
          & 19.7 & 17.6
          & \textbf{11.9} & \pmsd{0.5} \\
    \midrule
    \multirow{2}{*}{{\Qwen}$^{\ast}$}
      & \Defects & 0.030 & 0.040
          & \textbf{0.163} & \pmsd{0.055}
          & 0.045 & 0.100
          & \textbf{0.320} & \pmsd{0.028}
          & 3.2 & 12.6
          & 13.0 & \pmsd{1.5} \\
      & \GitBug & 0.015 & 0.010
          & \textbf{0.092} & \pmsd{0.006}
          & 0.065 & 0.098
          & \textbf{0.329} & \pmsd{0.020}
          & 10.0 & 5.5
          & 9.1 & \pmsd{0.6} \\
    \bottomrule
  \end{tabular}
  \end{adjustbox}

%% file: figures/scaffold-body.tex
  \pgfplotsset{
    scaffoldbody/.style={
      width=\linewidth,
      height=4.0cm,
      ymin=0,
      xmin=-0.3, xmax=1.3,
      bar width=9pt,
      tick label style={font=\scriptsize},
      label style={font=\scriptsize},
      axis line style={black!50},
      clip=false,
    },
  }

  \begin{tikzpicture}
    \begin{axis}[scaffoldbody,
        axis y line*=left,
        ymax=0.4,
        ytick={0,0.1,0.2,0.3,0.4},
        ylabel={\FoneReject{} $\uparrow$},
        xtick={0.05,0.95},
        x tick label style={font=\scriptsize, align=center},
        xticklabels={Patch acceptance\\(\SBL),
                     Fault localization\\(\Defects)},
        area legend,
        legend columns=-1,
        legend image post style={xscale=0.7},
        legend style={font=\tiny, draw=none, column sep=0.5ex,
                      at={(0.5,1.04)}, anchor=south},
        ymajorgrids,
        major grid style={black!12}]
      \addplot[ybar, bar shift=-10pt, fill=black!30]
        coordinates {(0.05,0.268)};
      \addplot[ybar, bar shift=0pt, fill=NavyBlue!50]
        coordinates {(0.05,0.301)};
      \addplot[ybar, bar shift=10pt, fill=NavyBlue]
        coordinates {(0.05,0.383)};
      \legend{\Acr, \AcrBody, \AgIntRef}
    \end{axis}
    \begin{axis}[scaffoldbody,
        axis y line*=right,
        axis x line=none,
        ymax=80,
        ytick={0,20,40,60,80},
        ylabel={\WastedAt{5} $\downarrow$}]
      \addplot[ybar, bar shift=-10pt, fill=black!30]
        coordinates {(0.95,71.0)};
      \addplot[ybar, bar shift=0pt, fill=NavyBlue!50]
        coordinates {(0.95,19.8)};
      \addplot[ybar, bar shift=10pt, fill=NavyBlue]
        coordinates {(0.95,17.2)};
    \end{axis}
  \end{tikzpicture}

%% file: tables/witness.tex
\begin{tabular}{@{}llrrr@{}}
  \toprule
  bug type & analyzer & \# & \AgInt{} & baseline \\
  \midrule
  \multirow{2}{*}{null dereference}
    & SonarQube & 197 & 119 (60.4\%) & 101 (51.3\%) \\
  \cmidrule(l){2-5}
    & Infer & 278 & 174 (62.6\%) & 152 (54.7\%) \\
  \midrule
  \multirow{2}{*}{resource leak}
    & SonarQube & 70 & 30 (42.9\%) & 27 (38.6\%) \\
  \cmidrule(l){2-5}
    & Infer & 47 & 32 (68.1\%) & 31 (66.0\%) \\
  \midrule
  \multicolumn{3}{@{}l}{turns/candidate}      & 34.7 & 37.0 \\
  \multicolumn{3}{@{}l}{executions/candidate} & 15.9 & 24.5 \\
  \multicolumn{3}{@{}l}{cost/candidate}       & \$3.66 & \$8.83 \\
  \bottomrule
\end{tabular}

%% file: tables/witness-soundness-floor.tex
\begin{tabular}{@{}crr@{}}
  \toprule
  real bug? & witnessed & not witnessed \\
  \midrule
  \cmark & 142 & 0 \\
  \xmark & 0   & 142 \\
  \bottomrule
\end{tabular}

%% file: src/conclusion.tex
\section{Conclusion}
\label{sec:conclusion}

We introduced \emph{Agentic Interpretation} (\AgInt), a framework in which the
harness governs how a software agent constructs and uses evidence. \AgInt{}
changes the role of the harness from a tool router into part of the agent's
reasoning process. The model proposes typed records, while the harness admits
and preserves them, exposes unresolved requirements, and licenses a verdict
only when task-defined obligations are met. The grounding lattice formalizes
how these records accumulate while leaving the model free to decide what to
investigate next.

Across four models and four labelled datasets, \AgInt{} generally improves
patch acceptance and fault localization over a plain agent and \Acr{}. The
improvements are most pronounced in settings where the plain agent struggles;
when \Opus{} already performs strongly, the methods are closer. The
matched-template control shows that \AgInt{} is not merely a better certificate
prompt: the persistent scaffold carries most of the fault-localization gain,
while the certificate body and gate drive patch-acceptance calibration. In
execution-backed bug witnessing, \AgInt{} raises null-dereference witness
yield, remains close to the shell agent on resource leaks, uses fewer
executions and incurs lower cost in the reported efficiency measurements, and
backs every positive with a harness-reproduced witness.

These results show that a harness can do more than route tools or enforce
guardrails. Once it defines what the agent's state is and how that state may
change, it gives the model durable structure to reason over and leaves a
certificate that a third party can recheck after the session ends. More
broadly, the correspondence between agent state and the lattices of program
analysis suggests that established analysis algorithms can guide the design of
future agent harnesses.

%% file: src/data_availability.tex
\section*{Data Availability}
Our data and artifacts are restricted by internal review. They will be made available after the paper gets published for artifact evaluation.